# Anisotropic Metamaterials Emulated by Tapered Waveguides: Application to Optical Cloaking

Igor I. Smolyaninov, Vera N. Smolyaninova[1], Alexander V. Kildishev[2], and Vladimir M. Shalaev[2]

[1]*Dept. of Physics Astronomy and Geosciences, Towson University, 8000 York Rd., Towson, MD 21252, USA*
[2]*Birck Nanotechnology Center, School of Electrical and Computer Engineering, Purdue University, West Lafayette, IN 47907, USA*

We demonstrate that metamaterial devices requiring anisotropic dielectric permittivity and magnetic permeability may be emulated by specially designed tapered waveguides. This approach leads to low-loss, broadband performance. Based on this technique, we demonstrate broadband electromagnetic cloaking in the visible frequency range on a scale ~100 times larger than the wavelength.

Current interest in electromagnetic metamaterials has been motivated to a large extent by the recent work on cloaking and transformation optics [1-7]. This interest has been followed by considerable effort aimed at the introduction of metamaterial structures that could be realized experimentally. Unfortunately, it appears difficult to develop metamaterials with low losses and broadband performance. The difficulties are especially severe in the visible frequency range, where obtaining good magnetic performance is challenging. Because of these problems, the existing experimental realizations of cloaking-like behavior are limited to very small sizes, on the scale of a few wavelengths only [8, 9]. In such cases, the shadow produced by an uncloaked object of the same size would not be strong anyway.

Here we demonstrate that metamaterial devices requiring anisotropic dielectric permittivity and magnetic permeability can be emulated by specially designed tapered waveguides. This approach leads to low-loss, broadband performance in the visible frequency range, which is difficult to achieve by other means. We apply this technique to electromagnetic cloaking and demonstrate broadband, two-dimensional electromagnetic cloaking in the visible frequency range on a scale roughly 100 times larger than that of the incident wavelength.

As a starting point, we show that the transformation optics approach allows us to map a planar region of space filled with an inhomogeneous, anisotropic metamaterial into an equivalent region of empty space with curvilinear boundaries (a tapered waveguide). We begin with Maxwell's curl equations

$$\tilde{\nabla} \times \tilde{\mathbf{H}} = -i\omega\tilde{\varepsilon}\tilde{\mathbf{E}}, \quad \tilde{\nabla} \times \tilde{\mathbf{E}} = i\omega\tilde{\mu}\tilde{\mathbf{H}}. \quad (1)$$

We use the following notations for the vector fields: $\tilde{\mathbf{E}} = \sum \tilde{e}_i \tilde{\mathbf{x}}_i$ and $\tilde{\mathbf{H}} = \sum \tilde{h}_i \tilde{\mathbf{x}}_i$, described in an orthogonal curvilinear system with unit vectors $\tilde{\mathbf{x}}_i$ and where the formal curl, $\tilde{\nabla} \times \tilde{\mathbf{v}}$, is given by $\tilde{\mathbf{x}}_1(\partial_2 \tilde{v}_3 - \partial_3 \tilde{v}_2)$, $\tilde{\mathbf{x}}_2(\partial_3 \tilde{v}_1 - \partial_1 \tilde{v}_3)$ and $\tilde{\mathbf{x}}_3(\partial_1 \tilde{v}_2 - \partial_2 \tilde{v}_1)$ [10].

The components of the actual physical fields $\mathbf{E} = \sum e_i \mathbf{x}_i$ and $\mathbf{H} = \sum h_i \mathbf{x}_i$ are connected to the fields in the material coordinates through $\tilde{e}_i = s_i e_i$ and $\tilde{h}_i = s_i h_i$ via the metric coefficients $s_i$. The tensors $\tilde{\varepsilon}$ and $\tilde{\mu}$ are given by $\mathbf{t}\varepsilon$ and $\mathbf{t}\mu$, with $\mathbf{t}$ being $\mathbf{t} = s_1 s_2 s_3 \mathbf{s}^{-2}$, and $\mathbf{s} = diag(s_1, s_2, s_3)$.

For comparing axisymmetric cloaking and imaging systems, it is convenient to match a given axisymmetric material domain to an equivalent inhomogeneous and anisotropic axisymmetric material distribution between two planes in the cylindrical coordinate system. In this case, the metric coefficients are $s_\rho = s_z = 1$, $s_\phi = \rho$. Therefore, $\rho e_\phi = \tilde{e}_\phi$, $\rho h_\phi = \tilde{h}_\phi$, and $e_i = \tilde{e}_i$, $h_i = \tilde{h}_i$ for $i = \rho, z$. Thus, any rotational coordinate $(\tilde{\rho}, \tilde{\phi}, \tilde{z})$ can be converted into the cylindrical format through

$$\tilde{\nabla} \times \tilde{\mathbf{H}} = -i\omega \tilde{\mathbf{d}}\tilde{\varepsilon}\tilde{\mathbf{E}}, \quad \tilde{\nabla} \times \tilde{\mathbf{E}} = i\omega \tilde{\mathbf{d}}\tilde{\varepsilon}\tilde{\mathbf{H}}, \quad (2)$$

(where $\tilde{\mathbf{d}} = diag(\tilde{\rho}, 1/\tilde{\rho}, \tilde{\rho})$) is equivalent to a cylinder coordinate system with material tensors $\tilde{\varepsilon}$ and $\tilde{\mu}$.

A direct application of the above formalism leads us to an experimental demonstration of optical cloaking as described below. We map an axisymmetric space between a spherical surface and a planar surface (Fig. 1(a)) onto a space between two planes (Fig. 1(b)), using the parametric description

$$\rho = \delta\sqrt{\beta\tilde{\rho}}/\tau, \quad z = \delta\tilde{z}/\tau, \quad \phi = \tilde{\phi}. \quad (3)$$

Here $\tau = \delta - \tilde{\rho} + \alpha$, $\alpha = \sqrt{\delta^2 + \tilde{z}^2}$, $\beta = 2\delta - \tilde{\rho}$, $\delta = \sqrt{z_0(2r_0 + z_0)} \approx \sqrt{2r_0 z_0}$, and $r_0$ and $z_0$ are the radius of the sphere and the minimal gap, respectively (Fig. 1a).

The transformation optics technique gives the diagonal components

$$\tilde{\varepsilon}_{\tilde{\rho}} = \delta\beta/(\alpha\tau), \quad \tilde{\varepsilon}_{\tilde{\phi}} = \delta^3/(\alpha\beta\tau), \quad \tilde{\varepsilon}_{\tilde{z}} = \alpha\delta/(\tilde{\rho}\tau) \quad (4)$$

of $\tilde{\mu} = \tilde{\varepsilon}$ tensors, distributed in an equivalent layer between two planes. Analysis of Eqs. (4) and Fig. 1(c) indicates that Eqs. (4) can be approximated with

$$\tilde{\varepsilon}_{\tilde{\rho}} = \tilde{\mu}_{\tilde{\rho}} \approx 1, \quad \tilde{\varepsilon}_{\phi} = \tilde{\mu}_{\phi} \approx (\delta/\beta)^2, \quad \tilde{\varepsilon}_{\tilde{z}} = \tilde{\mu}_{\tilde{z}} \approx \delta^2/(\beta\tilde{\rho}). \quad (5)$$

An ideal cloak should obey the following constrains [10]

$$\tilde{\varepsilon}_{\tilde{\rho}} = \rho\tilde{\rho}'/\tilde{\rho}, \quad \tilde{\varepsilon}_{\phi} = \tilde{\varepsilon}_{\tilde{\rho}}^{-1}, \quad \tilde{\varepsilon}_{\tilde{z}} = \rho/(\tilde{\rho}'\tilde{\rho}), \quad (6)$$

where $\tilde{\rho} = \tilde{\rho}(\rho)$ is a radial mapping function, $\tilde{\rho}' = d\tilde{\rho}/d\rho$, and $\tilde{\varepsilon} = \tilde{\mu}$. These conditions can be met if the refractive index ($n = \sqrt{\varepsilon}$) inside the gap between the sphere and the plane is chosen to be a simple radius-dependent function $n = \sqrt{\beta/\delta}$. In such a case we obtain

$$\tilde{\varepsilon}_{\tilde{\rho},i} \approx \beta/\delta, \quad \tilde{\varepsilon}_{\phi,i} = \delta/\beta, \quad \tilde{\varepsilon}_{\tilde{z},i} = \delta/\tilde{\rho}, \quad (7)$$

Note that the scale $\delta$ is chosen to avoid singularities $\delta > \max(\tilde{\rho})$, and the filling substance has an isotropic effective refractive index ranging from 2 to 1 for $\rho = [0, \delta]$. Since in general, $\tilde{\varepsilon}_{\tilde{\phi}}\tilde{\varepsilon}_{\tilde{z}} = (\rho/\tilde{\rho})^2$, and this ratio in (7) is equal to $\delta^2/(\beta\tilde{\rho})$, the mapping function yields $\rho(\tilde{\rho}) = \delta\sqrt{\tilde{\rho}/\beta}$, which is consistent with the approximated version of the initial mapping, $\rho = \delta\sqrt{\beta\tilde{\rho}}/(\delta - \tilde{\rho} + \alpha)$. The parameters of the spherical surface are as follows: radius of the sphere $r_0 = \delta^2/\tilde{z}_0$, and the minimal gap between the sphere and the plane is $z_0 = \delta^2/(r_0 + \alpha)$.

Figure 1(c) (solid lines) compares the behavior of the components of the effective anisotropic permittivity obtained in accordance with Eq. (5) with the corresponding components of the effective anisotropic permittivity (dashed lines) obtained in a waveguide filled with a material with a radius-dependent refractive index, $n = \sqrt{\beta/\delta}$. Thus, the dashed-line waveguide obeys the ideal cloak condition of Eq. (7). Note that the ratios of the two most important components ($\tilde{\varepsilon}_{\phi,i}, \tilde{\varepsilon}_{\tilde{z},i}$) and ($\tilde{\varepsilon}_{\phi}, \tilde{\varepsilon}_{\tilde{z}}$) are preserved both with and without the filling material; hence these components could exhibit similar cloaking-type dispersion under adiabatically changing parameters.

Equation (7) represents the *invisible body*, i.e., a self-cloaking arrangement. It is important to note that filling an initial domain between rotationally symmetric curvilinear boundaries, for example, with an anisotropic dielectric *allows for independent control over the effective magnetic and electric properties in the equivalent right-cylinder domain.*

In the semi-classical ray-optics approximation, the cloaking geometry may be simplified further for a family of rays with similar parameters. Our starting point is the

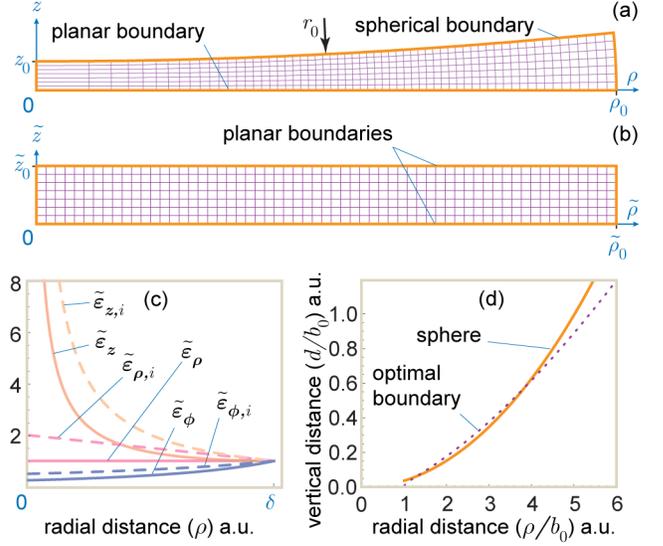

FIG. 1 (color). A space between a spherical surface and a planar surface (a) mapped onto a layer with planar boundaries (b). (c) Distribution of the radial, azimuthal, and axial (or vertical) diagonal components of permittivity and permeability in the equivalent planar waveguide. Dashed lines show the same components in the waveguide with a radius-dependent refractive index. (d) Normalized profile of the optimal waveguide shape plotted for a cloak radius of $b_0 = 172$ μm. The shape of the optimal waveguide may be approximated by a spherical surface placed on top of a flat surface, as shown by the dashed line.

semi-classical 2D cloaking Hamiltonian (dispersion law) introduced in [7] as $(\omega/c)^2 = k_\rho^2 + k_\phi^2(\rho - b)^{-2}$. Then we have the following equation

$$(\omega/c)^2 = k_\rho^2 + (k_\phi/\rho)^2 + b(k_\phi/\rho)^2(2\rho - b)/(\rho - b)^2. \quad (8)$$

For such a cylindrically symmetric Hamiltonian, the rays of light would flow without scattering around a cloaked region of radius $b$. Our aim is to "produce" the Hamiltonian of Eq. (8) in an optical waveguide (Fig. 2). Let us allow the thickness $d$ of the waveguide in the $z$-direction to change adiabatically with radius $\rho$. The top and bottom surfaces of the waveguide are coated with metal. The dispersion law for light in such a waveguide is given by

$$(\omega/c)^2 = k_\rho^2 + (k_\phi/\rho)^2 + [\pi l/d(\rho)]^2, \quad (9)$$

where $l$ is the transverse mode number. A photon launched into the $l^{th}$ mode of the waveguide stays in this mode as long as $d$ changes adiabatically [11]. Since the angular momentum of the photon $k_\phi = m$ is conserved, for each combination of $l$ and $m$ the cloaking Hamiltonian (8) can be emulated precisely by adiabatically changing $d(\rho)$.

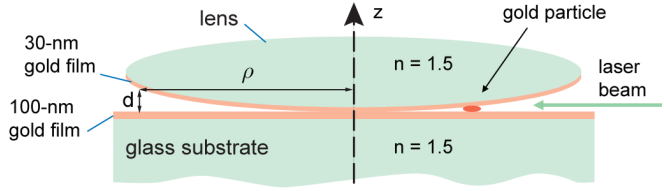

FIG. 2 (color). (a) Waveguide experiment for emulating the cloaking Hamiltonian.

A comparison of Eqs. (8) and (9) produces the following desired radial dependence of the waveguide thickness:

$$d = \pi \rho^{3/2} (l/m) \sqrt{(1 - b_{l,m}/\rho)/(2b_{l,m})}, \qquad (10)$$

where $b_{l,m}$ is the radius of the region that is "cloaked" for the photon launched into the $(l,m)$ mode of the waveguide.

The shape of such a waveguide is presented in Fig. 3(d). Thus, an electromagnetic cloaking experiment in a waveguide may be performed in a geometry that is identical to the classic geometry of the Newton rings observation [12], as shown in Fig. 2. An aspherical lens shaped according to Eq. (10) has to be used for a given $(l,m)$ mode cloak in the ideal cloak. This shape may be approximated by a spherical surface placed on top of a flat surface, shown by the solid line in Fig. 1(d).

Moreover, the waveguide geometry may be further improved to allow for cloaking in a multi-mode waveguide geometry. The choice of $b_{l,m} = b_0 (l/m)^2$ leads to the same desired shape of the waveguide for *all the* $(l,m)$ *modes* of a multi-mode waveguide, since in the leading order of $b_{l,m}/\rho$, $d = \pi \sqrt{\rho^3/(2b_0)}$. This equation describes the best-shaped aspherical lens for the observation of a broadband electromagnetic cloak.

In our experiments, a 4.5-mm diameter double convex glass lens (lens focus 6 mm) was coated on one side with a 30-nm gold film. The lens was placed with the gold-coated side down on top of a flat glass slide coated with a 70-nm gold film. The air gap between these surfaces has been used as an adiabatically changing waveguide. A simplified schematic of the test geometry is shown in Fig. 2 (not to scale). The point of contact between two gold-coated surfaces is visible in Fig. 3(a). Newton rings appear around the point of contact upon illumination of the waveguide with white light from the top. The radius of the $l^{th}$ ring is given by the expression, $\rho_p = \sqrt{(l + 1/2) r_0 \lambda}$, where $r_0$ is the lens radius. Figure 1(b) shows the central area around the point of contact; it appears bright since light reflected from the two gold-coated surfaces has the same phase.

Light from an argon ion laser was coupled to the waveguide via side illumination. Light propagation through the waveguide was imaged from the top using an optical microscope. Figures 3(e-h) show microscope images of the

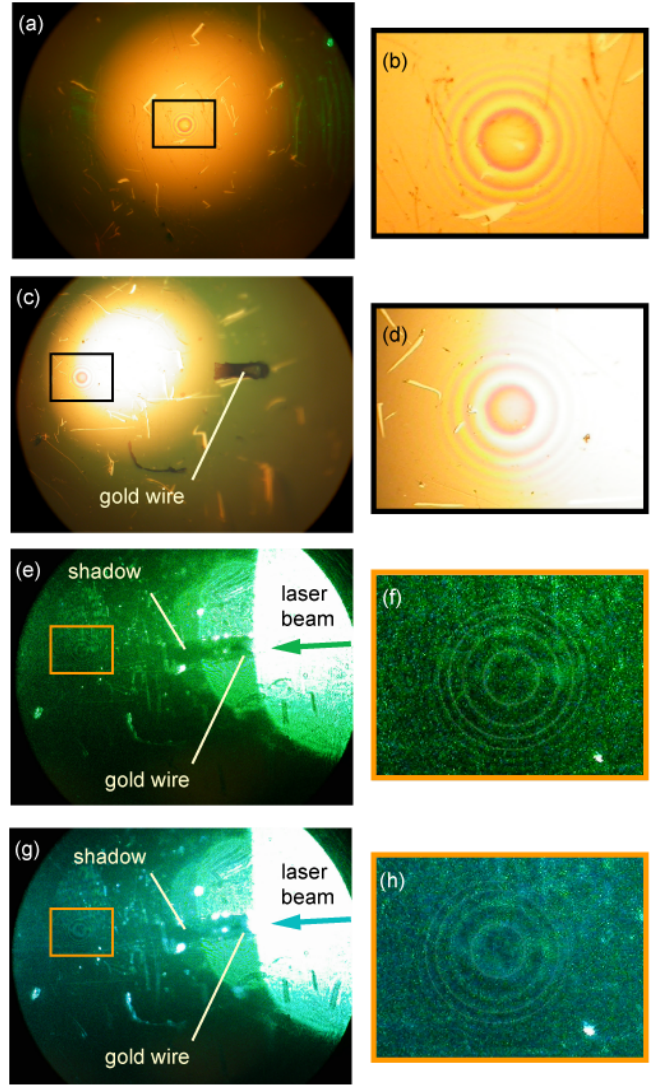

FIG. 3 (color). (a) Microscope image of the waveguide without a particle (white light coming from the top). (b) Magnified Newton rings taken from the frame in (a). (c) Image of the waveguide with a gold particle (white light coming from the top). (d) Magnified Newton rings taken from the frame in (c). A long shadow cast by the gold particle upon coupling (via side illumination) 515-nm (e) and 488-nm (g) laser light into the waveguide. Magnified images of the rings for 515-nm (f) and 488-nm (h) light taken from frames in (e) and (g).

light propagation through the waveguide in an experiment in which a gold particle cut from a 50-μm diameter gold wire is placed inside the waveguide. A pronounced long shadow is cast by the particle inside the waveguide (Fig. 3(e,g)). This is natural since the gold particle size is approximately equal to 100λ (note that the first dark Newton rings magnified in Figs. 3(f,h) have approximately the same size). Since the gold particle is located ~ 400 μm from the point of contact between the walls of the waveguide, the effective Hamiltonian around the gold particle differs strongly from the cloaking Hamiltonian of Eq. (8). Figure 3 represents the results of our best effort to insert a 150-μm long by 50-μm diameter gold particle inside the waveguide and orient it along the illumination direction. A few scratches, visible in Fig. 3, resulted from achieving this task.

While the gold particle casts a long and pronounced shadow, the area around the point of contact between the two gold-coated surfaces, which is roughly of the same size as the particle, casts no shadow at all (Fig. 3(e-g)). This is an observation which would be extremely surprising in the absence of the theoretical description presented above. For the $l$-th mode in the waveguide of Fig. 2, the cut-off radius is given by the same expression as that of the radius of the $l$-th Newton ring; thus, no photon launched into the waveguide can reach an area within the radius $\rho_0 = \sqrt{\lambda r_0/2} \approx 30\,\mu\text{m}$ from the point of contact between the two gold-coated surfaces. This is consistent with the fact that the area around the point of contact appears dark in Fig. 4. Even though photons may couple to surface plasmons [13] near the cut-off point of the waveguide, the propagation length of the surface plasmons at 515 nm is only few micrometers. Thus, the area around the point of contact about 50 μm in diameter is about as opaque for guided photons as the ~50-μm gold particle from Fig. 3(e,g), which casts a pronounced shadow. Nevertheless, there appears to be no shadow behind the cut-off area of the waveguide (Fig. 3(e-h)). The observed cloaking behavior appears to be broadband, which is consistent with the theory presented above: the images in Fig. 3(e-h) were taken at 488 nm (Fig. 3(g,h)) and 515 nm (Fig. 3(e,f)).

To test this waveguide cloaking independently of any other scatterers in the gap, which generally could interfere with the clean observation of the cloaking effect, separate images of the point of contact have been made with no wire in the waveguide, as shown Fig. 4(a,b). The images in Fig. 4 have been also taken at laser lines of 488 nm (Fig. 4(a)) and 515 nm (Fig. 4(b)). Along with the images of Fig. 3(e-h), the magnified no-shade rings of Fig. 4 give yet another indication that broadband cloaking performance is observed.

It is also interesting to note that the geometry of our cloaking experiment is similar to the geometry recently proposed in [3], where broadband cloaking has been also predicted.

We conclude that metamaterial devices requiring anisotropic dielectric permittivity and magnetic permeability may be emulated by specially designed tapered waveguides. We note that filling the specifically shaped waveguide domain with an anisotropic dielectric allows for independent control over the effective magnetic and electric properties in the equivalent right-cylinder domain. A similar methodology for waveguide-confined light has been used in a different application context in [14,15]. We also show that this approach leads to low-loss, broadband performance for cloaking applications.

In summary, by using an axisymmetric waveguide between a planar gold film and a gold-coated spherical lens, we experimentally demonstrated a broadband optical cloaking effect in the visible frequency range for an object, which is two orders of magnitude larger than the wavelength of the incident light.

This work was supported in part by ARO-MURI award 50342-PH-MUR and by the NSF grant DMR-0348939.

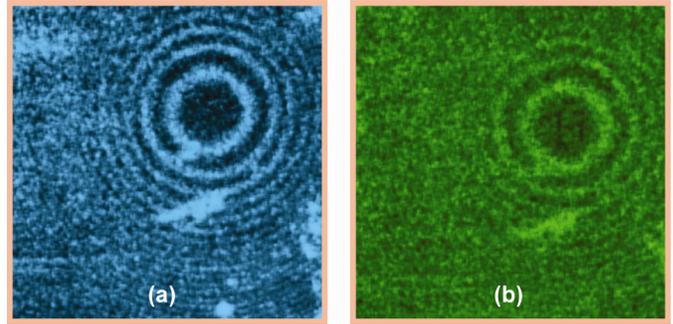

FIG. 4 (color). Magnified images of the rings for (a) 488-nm and (b) 515-nm laser illumination. In both cases, no wire is placed in the waveguide.